\documentclass[preprint,12pt]{elsarticle}
 \usepackage{graphicx}

\begin{document}

\begin{frontmatter}
\title{Anomalies and chiral symmetry in QCD} 
\author{Michael Creutz}
\address{Physics Department, Brookhaven National Laboratory\\
Upton, NY 11973, USA
}
\begin{abstract}
{I review some aspects of the interplay between anomalies and
 chiral symmetry.  The quantum anomaly that breaks the U(1) axial
 symmetry of massless QCD leaves behind a flavor-singlet discrete
 chiral invariance.  When the mass is turned on this residual symmetry
 has a close connection with the strong CP violating parameter theta.
 One result is that a first order transition is usually expected when
 the strong CP violating angle passes through pi.  This symmetry can
 be understood either in terms of effective chiral Lagrangians or in
 terms of the underlying quark fields.}
\end{abstract}


\end{frontmatter}

\section{Introduction}

The classical Lagrangian for QCD couples left and right handed quark
fields only through mass terms.  Thus naively the massless theory will
have independent conserved currents associated with each handedness.
For $N_f$ massless flavors, this would be an independent $U(N_f)$
symmetry associated with each chirality, giving a full symmetry that
is often written in terms of axial and vector fields as
$U(N_f)_V\times U(N_f)_A$.  As is well known, this full symmetry does
not survive quantization, being broken to a $SU(N_f)_V\times
SU(N_f)_A\times U(1)_B$, where the $U(1)_B$ represents the symmetry of
baryon number conservation.  The only surviving axial symmetries of
the quantum theory are non-singlet under flavor symmetry.

This breaking of the classical $U(1)$ axial symmetry is tied to the
possibility of introducing into massive QCD a CP violating parameter,
usually called $\Theta$.  For an extensive recent review of this
quantity, see Ref.~\cite{Vicari:2008jw}.  While such a term is allowed
from fundamental principles, experimentally it appears to be extremely
small.  This raises an unresolved puzzle for attempts to unify the
strong interactions with the weak.  Since the weak interactions do
violate CP, why is there no residue of this remaining in the strong
sector below the unification scale?

One goal here is to provide a qualitative picture of the $\Theta$
parameter in meson physics.  I will concentrate on symmetry alone and
will not attempt to rely on any specific form for an effective
Lagrangian.  I build on a connection between $\Theta$ and a
flavor-singlet $Z_{N_f}$ symmetry that survives the anomaly.  This
symmetry predicts that, if the lightest quarks are massive and
degenerate, then a first order transition is expected when $\Theta$
passes through $\pi$.  This transition is quite generic, and only can
be avoided under limited conditions with one quark considerably
lighter than the others.  It will also become clear why the sign of
the quark mass is relevant for an odd number of flavors, an effect
unseen in naive perturbation theory.

This picture has evolved over many years.  The possibility of the
spontaneous CP violation occurring at $\Theta=\pi$ is tied to what is
known as Dashen's phenomenon \cite{dashen}, first noted even before
the days of QCD.  In the mid 1970's, 't Hooft \cite{'tHooft:fv}
elucidated the underlying connection between the chiral anomaly and
the topology of gauge fields.  Later Witten \cite{Witten:1980sp} used
large gauge group ideas to discuss the behavior at $\Theta=\pi$ in
terms of effective Lagrangians.  Ref.~\cite{oldeffective} lists a few
of the early studies of the effects of $\Theta$ on effective
Lagrangians via a mixing between quark and gluonic operators.  The
topic continues to appear in various contexts; for example,
Ref.~\cite{Boer:2008ct} contains a different approach to understanding
the transition at $\Theta=\pi$ via the framework of the two-flavor
Nambu Jona-Lasinio model.

I became interested in the issues while trying to understand
difficulties with formulating chiral symmetry on the lattice.  Much of
the picture presented here is implicit in my 1995 paper on quark
masses \cite{Creutz:1995wf}.  Since then the topic has become highly
controversial, with the realization of ambiguities precluding a
vanishing up quark mass solving the strong CP problem
\cite{Creutz:2003xc} and the appearance of an inconsistency with one
of the popular algorithms in lattice gauge theory
\cite{Creutz:2008nk}.  Despite the controversies, both are direct
consequences of the interplay of the anomaly and chiral symmetry
discussed here.  The fact that these issues remain so disputed is what
has driven me to write this overview.

Section \ref{effective} reviews the conventional picture of
spontaneous chiral symmetry breaking wherein the light pseudoscalars
are identified as approximate Goldstone bosons in an effective
Lagrangian.  Here I introduce a mass contribution, coming from the
anomaly, for for the flavor singlet pseudoscalar meson.  Section
\ref{quarks} reviews how the anomaly arises in terms of the underlying
quark fields and gauge fields with non-trivial topology.  Here the
fact that the theory requires a regulator that breaks chiral symmetry
is crucial.  Section \ref{znf} returns to the effective field picture
and exposes the flavor singlet $Z_{N_f}$ symmetry.  In Section
\ref{mass} I add in a small quark mass to break the chiral symmetry.
Depending on the mass and $N_f$, the effective potential can display
multiple meta-stable minima.  Doing an anomalous chiral rotation on
the mass term brings in the parameter $\Theta$.  The first order
transition at $\Theta=\pi$ corresponds to a jump of the physical
vacuum between two distinct degenerate minima.  At this point the
theory spontaneously breaks CP invariance.  Section
\ref{nondegenerate} discusses increasing the mass of one quark species
to allow an interpolation between various $N_f$.  Section
\ref{summary} draws concluding remarks.
 
I start with a few reasonably uncontroversial assumptions.  First QCD
with $N_f$ light quarks should exist as a field theory and exhibit
confinement in the usual way.  I assume the validity of the standard
picture of chiral symmetry breaking involving a quark condensate
$\langle\overline\psi\psi\rangle\ne 0$.  The conventional chiral
perturbation theory based on expanding in masses and momenta around
the chiral limit should make sense.  I assume the usual result that
the anomaly generates a mass for the $\eta^\prime$ particle with this
mass surviving in the chiral limit.  And I consider $N_f$ small enough
to avoid any potential conformal phase of QCD \cite{Banks:1981nn}.

Throughout I use the language of continuum field theory.  I do have in
mind that some non-perturbative regulator has been imposed to define
various products of fields, such as the condensing combination
$\sigma=\overline\psi\psi$.  For a momentum space cutoff, I assume
that it is much larger than $\Lambda_{QCD}$.  Correspondingly, for a
lattice cutoff, then I imagine that the lattice spacing is much
smaller than $1/\Lambda_{QCD}$.  Thus I ignore any lattice artifacts
that are expected to vanish in the continuum limit.

\section {Effective potentials}
\label{effective}

I begin by considering an effective potential $V$ as a function of the
various meson fields in the problem.  Intuitively, $V$ represents the
energy of the lowest state for a given field expectation.  More
formally, this can be defined in the standard way via a Legendre
transformation.  I will ignore the well known result that effective
potentials must be convex functions of their arguments.  This is
easily understood in terms of a Maxwell construction involving the
phase separation that will occur if one asks for a field expectation
in what would otherwise be a concave region.  Ignoring this effect
allows me to use the normal language of spontaneous symmetry breaking
corresponding to having an effective potential with more than one
minimum.  When the underlying theory possesses some symmetry but the
individual minima do not, spontaneous breaking comes about when the
vacuum selects one minimum arbitrarily.

I frame the discussion in terms of composite scalar and pseudoscalar fields
\begin{equation}
\matrix{
\sigma&\sim&\overline\psi \psi\cr
\pi_\alpha&\sim& i\overline\psi \lambda_\alpha \gamma_5 \psi \cr
\eta^\prime &\sim& i\overline\psi\gamma_5\psi \cr
\delta_\alpha&\sim& \overline\psi \lambda_\alpha \psi\cr
}
\end{equation}
Here the $\lambda_\alpha$ are the generalization of the usual
Gell-Mann matrices to $SU(N_f)$.  The $\delta$ field is listed here
for completeness, but will not play a role in the following
discussion.  As mentioned earlier, I assume some sort of regulator,
perhaps a lattice, is in place to define these products of fields at
the same point.

Initially consider degenerate quarks with a small common mass $m$.  I
also begin by restricting $N_f$ to be even, returning later to the
subtleties arising for an odd number of flavors.  And, as mentioned
earlier, I keep $N_f$ small enough to maintain asymptotic freedom as
well as to avoid any possible conformal phases.

\begin{figure*}
\centering
\includegraphics[width=2.5in]{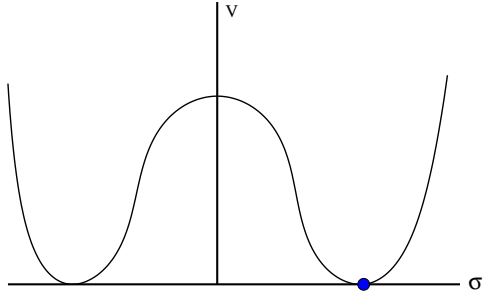}
\caption{\label{v1} 
Spontaneous chiral symmetry breaking is represented by a double well
effective potential with the vacuum settling into one of two possible
minima.  In this minimum chiral symmetry is broken by the selection of
a specific value for the quark condensate. 
}
\end{figure*}

The conventional picture of spontaneous chiral symmetry breaking at
$m=0$ begins with the vacuum acquiring a quark condensate with
$\langle\overline\psi\psi\rangle =\langle\sigma\rangle =v \ne 0$.  In
terms of the effective potential, $V(\sigma)$ should acquire a double
well structure, as sketched in Fig.~\ref{v1}.  The symmetry under
$\sigma\leftrightarrow -\sigma$ is associated with the invariance of
the action under a flavored chiral rotation.  For example, with two
flavors the change of variables
\begin{equation}
\matrix{
\psi\rightarrow e^{i\pi\tau_3\gamma_5/2}\psi\cr
\overline \psi\rightarrow \overline\psi e^{i\pi\tau_3\gamma_5/2}\cr
}
\end{equation}
leaves the massless action invariant but changes the sign of $\sigma$.
Here $\tau_3$ is the conventional Pauli matrix corresponding to the
third component of isospin.

\begin{figure*}
\centering
\includegraphics[width=2.5in]{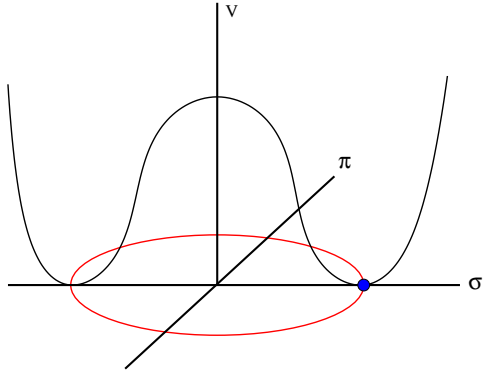}
\caption{\label{v0} 
The flavor non-singlet pseudoscalar mesons are Goldstone bosons
corresponding to flat directions  in the effective potential.
}
\end{figure*}

Extending the effective potential to a function of the non-singlet
pseudoscalar fields gives the standard picture of Goldstone bosons.
These are massless when the quark mass vanishes, corresponding to
$N_f^2-1$ ``flat'' directions for the potential.  It is useful to
introduce the
$SU(N_f)$ valued effective field
\begin{equation}
\Sigma=e^{i\lambda_\alpha\pi_\alpha/F_\pi}\sim
\overline\psi_L\psi_R
\end{equation}
where the matrices $\lambda$ generalize of the Gell-mann matrices to
$SU(N_f)$, flavor indices that make this a matrix quantity are
suppressed, and the pion decay constant $F_\pi$ is inserted as a
convenient normalization but will play no role in the qualitative
discussion here.  The left and right Fermi fields are defined as usual
by $\psi_{L,R}={1\mp \gamma_5 \over 2}\psi$.  In terms of $\Sigma$ the
chiral invariance of the potential takes the simple form
\begin{equation}
V(\Sigma)=V(g_L^\dagger \Sigma g_R)
\end{equation}
where $g_L$ and $g_R$ are arbitrary elements of $SU(N_f)$.  One flat
direction is sketched in Fig.~\ref{v0}.

\begin{figure*}
\centering
\includegraphics[width=2.5in]{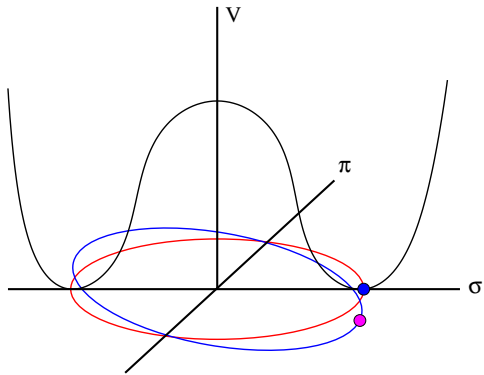}
\caption{\label{v3}
A small quark mass term tilts the effective potential, selecting one
direction for the true vacuum and giving the Goldstone bosons a mass.
}
\end{figure*}

The introduction of a small mass for the quarks effectively tilts the
potential $V(\sigma)\rightarrow V(\sigma)-m\sigma$.  This selects one
of the minima as the true vacuum, driving the above matrix $\Sigma
\rightarrow I$.  The tilting of the potential breaks the global
symmetry and gives the Goldstone bosons a small mass proportional to
the square root of the quark mass, as sketched in Fig.~\ref{v3}

This picture is, of course, completely standard.  It is also common
lore that the anomaly prevents the $\eta^\prime$ from being a
Goldstone boson and leaves it with a mass of order $\Lambda_{QCD}$
even in the massless quark limit.  The effective potential $V$ must
not be symmetric under the following anomalous rotation by an angle
$\phi$
\begin{equation}
\label{rotate}
\matrix{
\sigma\rightarrow\phantom{+}  \sigma \cos(\phi)+\eta^\prime \sin(\phi)\cr
\eta^\prime\rightarrow -\sigma \sin(\phi)+\eta^\prime \cos(\phi).\cr
}
\end{equation}

If we consider the effective potential as a function of the fields
$\sigma$ and $\eta^\prime$, it should have a minimum at $\sigma\sim v$
and $\eta^\prime \sim 0$.  Expanding about that point we expect a
qualitative form
\begin{equation}
V(\sigma,\eta^\prime)\sim m_\sigma^2 (\sigma-v)^2 
+ m_{\eta^\prime}^2 {\eta^\prime}^2
+O((\sigma-v)^3, {\eta^\prime}^4)
\end{equation}
where both $m_\sigma$ and $m_{\eta^\prime}$ remain of order
$\Lambda_{QCD}$, even in the chiral limit.  And, at least with an even
number of flavors as considered here, there should be a second minimum
with $\sigma\sim -v$.  Expanding about this point gives the 
\begin{equation}
V(\sigma,\eta^\prime)\sim m_\sigma^2 (\sigma+v)^2 
+ m_{\eta^\prime}^2 {\eta^\prime}^2
+O((\sigma+v)^3, {\eta^\prime}^4).
\end{equation}

At this point one can ask whether we know anything else about the
effective potential in this $(\sigma,\eta^\prime)$ plane.  In section
\ref{znf} I show that indeed we do, and the potential has a total of
$N_f$ equivalent minima in the chiral limit.  But first I digress
to review how the above minima arise in quark language.

\section{Quark fields}
\label{quarks}

The classical QCD Lagrangian has a symmetry under a rotation of the
underlying quark fields
\begin{equation}
\matrix{
\psi \rightarrow e^{i\phi\gamma_5/2}\psi\cr
\overline\psi \rightarrow \overline\psi e^{i\phi\gamma_5/2}\cr
}
\end{equation}
This corresponds directly to the transformation of the composite
fields given in Eq.~\ref{rotate}.  This symmetry is ``anomalous'' and
thus any regulator must break it with a remnant surviving in the
continuum limit.  The specifics of how this works depend on the
details of the regulator, but a simple understanding
\cite{Fujikawa:1979ay} comes from considering the fermionic measure in
the path integral.  If we make the above rotation on the field $\psi$,
the measure changes by the determinant of the rotation matrix
\begin{equation}
\label{measure}
d\psi\rightarrow |e^{-i\phi\gamma_5/2}| d\psi
=e^{-i\phi {\rm Tr}\gamma_5/2} d\psi.
\end{equation}
Here the subtlety of the regulator comes in.  Naively $\gamma_5$ is a
simple four by four traceless matrix.  If it is indeed traceless, then
the measure would be invariant.  However in the regulated theory
this is not the case.  This is intimately tied with the index theorem
for the Dirac operator in topologically non-trivial gauge fields.

A typical Dirac action takes the form $\overline\psi (D+m)\psi$ with
$D$ a function of the gauge fields.  In the naive continuum theory $D$
is anti-Hermitian, $D^\dagger=-D$, and anti-commutes with $\gamma_5$,
i.e. $[D,\gamma_5]_+=0$.  What complicates the issue with fermions is
the well known index theorem: if a background gauge field has winding
$\nu$, then there are known to be at least $\nu$ exact zero
eigenvalues of $D$.  Furthermore, on the space spanned by the
corresponding eigenvectors, $\gamma_5$ can be simultaneously
diagonalized with $D$.  The net winding number equals the number of
positive eigenvalues of $\gamma_5$ minus the number of negative
eigenvalues.  This theorem is well known and well reviewed elsewhere
\cite{Coleman:1978ae}.  Here I only will use the fact that in this
subspace the trace of $\gamma_5$ does not vanish, but equals $\nu$.

What about the higher eigenvalues of $D$?  Because $[D,\gamma_5]_+=0$,
these appear in opposite sign pairs; i.e. if $D|\psi\rangle
=\lambda|\psi\rangle$ then $D\gamma_5|\psi\rangle
=-\lambda\gamma_5|\psi\rangle$.  For an anti-Hermitean $D$, these modes
are orthogonal with $\langle\psi|\gamma_5\psi\rangle=0$.  As a
consequence, $\gamma_5$ is traceless on the subspace spanned by each
pair of eigenvectors.  

So what happened to the opposite chirality states to the zero modes?
In a regulated theory they are in some sense ``above the cutoff.''  In
a simple continuum discussion they have been ``lost at infinity.''
With a lattice regulator there is no ``infinity''; so, something more
subtle must happen.  With the overlap\cite{Ginsparg:1981bj} or
Wilson\cite{Wilson:1975id} fermions, one gives up the anti-Hermiticity
of $D$.  Most eigenvalues still occur in conjugate pairs and do not
contribute to the trace of $\gamma_5$.  However, in addition to the
small real eigenvalues representing the zero modes, there are
additional modes where the eigenvalues are large and real.  With
Wilson fermions these appear as massive doubler states.  With the
overlap, the eigenvalues are constrained to lie on a circle.  In this
case, for every exact zero mode there is another mode of opposite
chirality lying on the opposite side of the circle.  These modes are
effectively massive and break chiral symmetry.

So with the regulator in place, the trace of $\gamma_5$ does not
vanish on gauge configurations of non-trivial topology.  The change of
variables indicated in Eq.~\ref{measure} introduces into the path
integral a modification of the weighting by a factor
\begin{equation}
e^{-i\phi {\rm Tr}\gamma_5}=e^{-i\phi N_f\nu}
\end{equation}
here I consider applying the rotation to all flavors equally, thus the
factor of $N_f$ in the exponent.  The conclusion is that gauge
configurations that have non-trivial topology receive a complex weight
after the anomalous rotation.  This changes the underlying physics and
gives an inequivalent theory.  Although not the topic of discussion
here, note that this factor introduces a sign problem if one wishes to
study this physics via Monte Carlo simulations.  Here I have treated
all $N_f$ flavors equivalently; this corresponds to dividing the
conventionally defined CP violation angle, to be discussed later,
equally among the flavors, i.e. effectively $\phi=\Theta/N_f$.

The necessary involvement of both small and large eigenvalues warns of
the implicit danger in attempts to separate infrared from ultraviolet
effects.  When the anomaly is concerned, going to short distances is
not sufficient for ignoring non-perturbative effects related to
topology.

\section{A discrete chiral symmetry}
\label{znf}

\begin{figure*}
\centering
\includegraphics[width=2.5in]{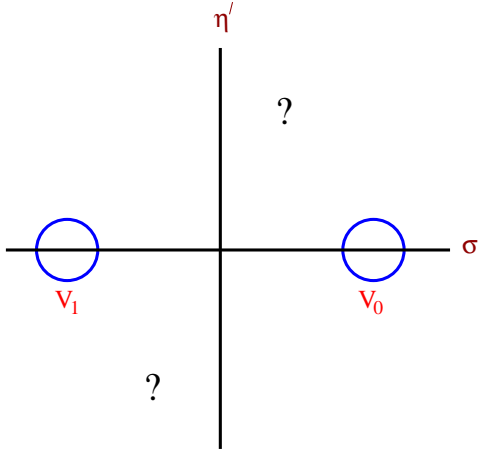}
\caption{\label{potential0}
We have two minima in the  $\sigma,\eta^\prime$ plane located at
$\sigma=\pm v$ and $\eta^\prime=0$.  Can we find any other minima?
}
\end{figure*}

I now return to the effective Lagrangian language of before.  For the
massless theory, the symmetry under $\sigma\leftrightarrow -\sigma$
indicates that we expect at least two minima for the effective
potential considered in the $\sigma,\eta^\prime$ plane.  These are
located as sketched in Fig.~\ref{potential0}.  Do we know anything
about the potential elsewhere in this plane?  The answer is yes,
indeed there are actually $N_f$ equivalent minima.

\begin{figure*}
\centering
\includegraphics[width=2.5in]{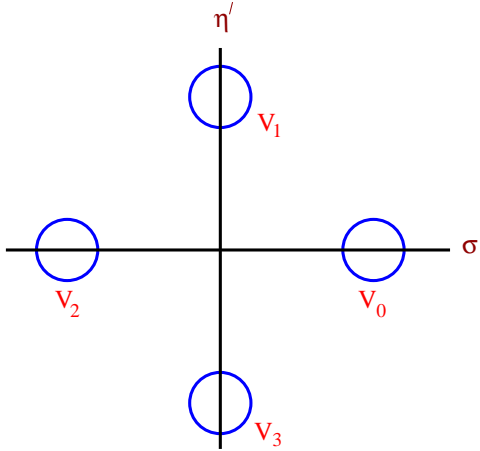}
\caption{\label{potential1}
For four flavors we have four equivalent
minima in the  $\sigma,\eta^\prime$ plane.  This generalizes to $N_f$
minima with $N_f$ flavors.
}
\end{figure*}

As noted before, due to the anomaly the singlet rotation
\begin{equation}
\psi_L\rightarrow e^{i\phi}\psi_L
\label{singlet}
\end{equation}
is not a valid symmetry of the theory for generic values of the angle
$\phi$.  On the other hand, flavored chiral symmetries should
survive, and in particular 
\begin{equation}
\psi_L\rightarrow
g_L\psi_L = e^{i\phi_\alpha \lambda_\alpha}\psi_L
\label{flavored}
\end{equation}
should be a valid symmetry for any set of angles $\phi_\alpha$.  The
point of this section is that, for special special discrete values of
the angles, the rotations in Eq.~\ref{singlet} and Eq.~\ref{flavored}
can coincide.  At such values the singlet rotation is a valid
symmetry.  In particular, note that
\begin{equation}
g=e^{2\pi i\phi/N_f} \in Z_{N_f} \subset SU(N_f).
\end{equation}
Thus a valid discrete symmetry involving only $\sigma$ and
$\eta^\prime$ is
\begin{equation}
\matrix{
\sigma\rightarrow \phantom{+} \sigma\cos(2\pi/N_f)
+\eta^\prime \sin(2\pi/N_f)\cr
\eta^\prime\rightarrow -\sigma \sin(2\pi/N_f)+\eta^\prime
\cos(2\pi/N_f).\cr
}
\end{equation}
The potential $V(\sigma,\eta^\prime)$ has a $Z_{N_f}$ symmetry
manifested in $N_f$ equivalent minima in the $(\sigma,\eta^\prime)$
plane.  For four flavors this structure is sketched in
Fig.~\ref{potential1}.

This discrete flavor singlet symmetry arises from the trivial fact
that $Z_N$ is a subgroup of both $SU(N)$ and $U(1)$.  At the quark
level the symmetry is easily understood since the 't Hooft vertex,
responsible for the chiral anomaly, receives one factor from every
flavor.  With $N_F$ flavors, these multiply together making
\begin{equation}
\psi_L\rightarrow e^{2\pi i/N_f} \psi_L
\end{equation}
a valid symmetry even though rotations by smaller angles are not.

\begin{figure*}
\centering
\includegraphics[width=2.5in]{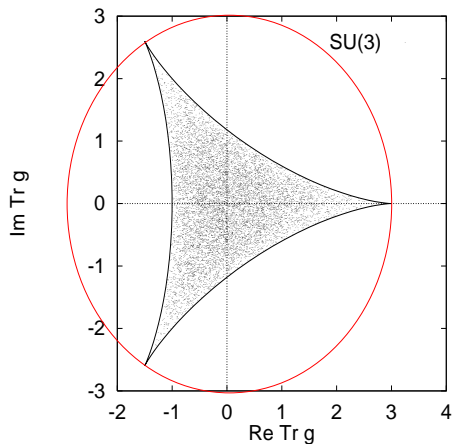}
\caption{\label{su3circle} The real and imaginary parts for the traces
of 10,000 randomly chosen $SU(3)$ matrices.  All points lie within the
boundary representing matrices of the form $\exp(i\phi\lambda_8)$.
The tips of the three points represent the center of the group.  The
outer curve represents the boundary that would be found if the group
was the full $U(1)$.  Taken from Ref.~\cite{Creutz:1995wf}.  }
\end{figure*}

\begin{figure*}
\centering
\includegraphics[width=2.5in]{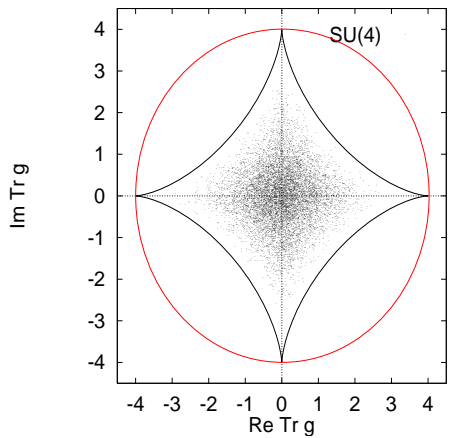}
\caption{\label{su4circle} The generalization of Fig.~\ref{su3circle}
to $SU(4)$.  The real and imaginary parts for the traces of 10,000
randomly chosen $SU(4)$ matrices.  Taken from
Ref.~\cite{Creutz:1995wf}.  }
\end{figure*}

The role of the $Z_N$ center of $SU(N)$ is illustrated graphically in
Fig.~\ref{su3circle}, taken from Ref.~\cite{Creutz:1995wf}.  Here I plot
the real and the imaginary parts of the traces of 10,000 $SU(3)$
matrices drawn randomly with the invariant group measure.  The region
of support only touches the $U(1)$ circle at the elements of the
center.  All elements lie on or within the curve mapped out by
elements of form $\exp(i\phi\lambda_8)$.  Fig.~\ref{su4circle} is a
similar plot for the group $SU(4)$.

\section{Massive quarks}
\label{mass}

As discussed earlier and illustrated in Fig.~\ref{potential1}, a quark
mass term $-m\overline\psi\psi\sim -m\sigma$ is represented by a
``tilting'' of the effective potential.  This selects one of the
minima in the $\sigma,\eta^\prime$ plane as the true vacuum.  For
masses small compared to the scale of QCD, the other minima will
persist, although due to the flat flavor non-singlet directions, some
of them will become unstable under small fluctuations.  Counting the
minima sequentially with the true vacuum having $n=0$, each is
associated with small excitations in the pseudo-Goldstone directions
having an effective mass of $m_\pi^2 \sim m \cos(2\pi n/N_f)$.
Thus when $N_f$ exceeds four, there will be more than one meta-stable
state.

\subsection{Twisted tilting}

Conventionally the mass tilts the potential downward in the $\sigma$
direction.  However, it is interesting to consider tilts in other
directions in the $\sigma,\eta^\prime$ plane.  This can be
accomplished by doing an anomalous rotation on the mass term
\begin{equation}\matrix{
-m\overline\psi\psi&\rightarrow&
-m\cos(\phi)\overline\psi\psi
-im\sin(\phi)\overline\psi\gamma_5\psi\cr
&\sim& -m\cos(\phi)\sigma-m\sin(\phi)\eta^\prime\cr
}
\end{equation}
Were it not for the anomaly, this would just be a redefinition of
fields.  However the same effect that gives the $\eta^\prime$ its mass
indicates that this new form for the mass term gives an inequivalent
theory.  As $i\overline\psi\gamma_5\psi$ is odd under CP, this theory
is explicitly CP violating.

\begin{figure*}
\centering
\includegraphics[width=2.5in]{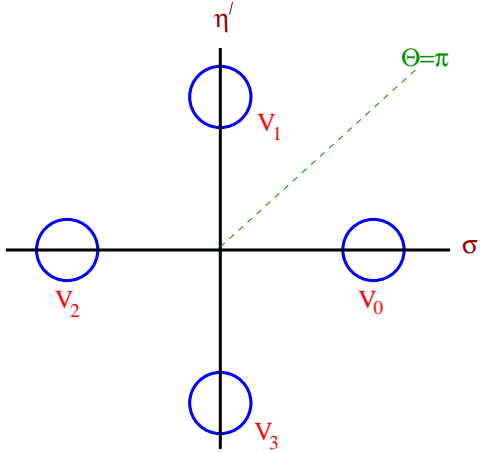}
\caption{\label{potential}
With massive quarks and a twisting angle of $\phi=\pi/N_f$, two of the
minima in the $\sigma,\eta^\prime$ plane become degenerate.  This
corresponds to a first order transition at $\Theta=\pi$.
}
\end{figure*}

The conventional notation for this effect involves the angle
$\Theta=N_f\phi$.  Then the $Z_{N_f}$ symmetry amounts to a $2\pi$
periodicity in $\Theta$.  As Fig.~\ref{potential} indicates, at
special values of the twisting angle $\phi$, there will exist two
degenerate minima.  This occurs, for example, at $\phi=\pi/N_f$ or
$\Theta=\pi$.  As the twisting increases through this point, there
will be a first order transition as the true vacuum jumps from one
minimum to the next.

\subsection{Odd $N_f$}

\begin{figure*}
\centering
\includegraphics[width=2.5in]{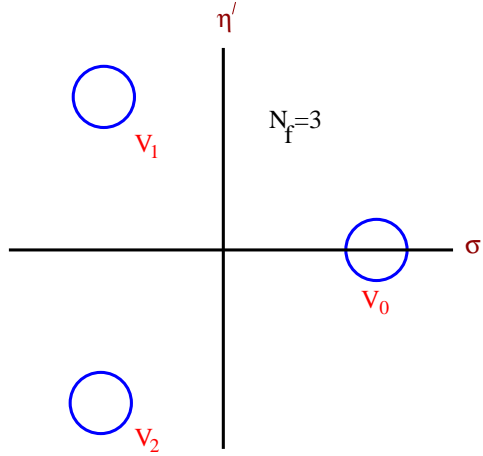}
\caption{\label{potential2}
For odd $N_f$, such as the $SU(3)$ case sketched here, QCD is not
symmetric under changing the sign of the quark mass.
Negative mass corresponds to taking $\Theta=\pi$.
}
\end{figure*}

One interesting consequence of this analysis is the behavior of QCD
with an odd number of flavors.  The group $SU(N_f)$ with odd $N_f$
does not include the element $-1$.  In particular, the $Z_{N_f}$
structure is not symmetric under reflections about the $\eta^\prime$
axis.  Fig.~\ref{potential2} sketches the situation for $SU(3)$.  One
immediate consequence is that positive and negative mass are not
equivalent.  Indeed, a negative mass corresponds to $\Theta=\pi$ where
a spontaneous breaking of CP is expected.  In this case the simple
picture sketched in Fig.~\ref{v1} no longer applies.

At $\Theta=\pi$ the theory lies on top of a first order phase
transition line.  A simple order parameter for this
transition is the expectation value for the $\eta^\prime$ field.
As this field is odd under CP symmetry, this is another way to see
that negative mass QCD with an odd
number of flavors spontaneously breaks CP.  This does not contradict
the Vafa-Witten theorem \cite{Vafa:1984xg} because in this
regime the fermion determinant is not positive definite.

Note that the asymmetry in the sign of the quark mass is not easily
seen in perturbation theory.  Any quark loop in a perturbative diagram
can have the sign of the quark mass flipped by a $\gamma_5$
transformation.  It is only through the subtleties of regulating the
divergent triangle diagram \cite{Adler:gk}, \cite{Bell:ts} that the
sign of the mass enters.

\begin{figure*}
\centering
\includegraphics[width=2.5in]{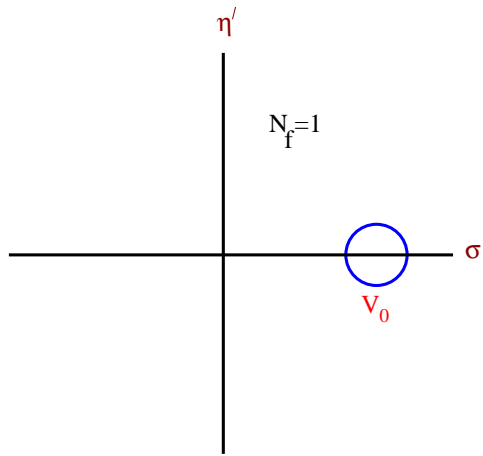}
\caption{\label{potential3} The effective potential for one-flavor QCD
with small quark mass has a unique minimum in the $\sigma,\eta^\prime$
plane.  The minimum is shifted from zero due to the effect of the 't
Hooft vertex.  }
\end{figure*}

A special case of an odd number of flavors is one-flavor QCD.  In this
case the anomaly removes all chiral symmetry and there is a unique
minimum in the $\sigma,\eta^\prime$ plane, as sketched in
Fig.~\ref{potential3}.  This minimum does not occur at the origin,
being shifted to $\langle \overline\psi \psi\rangle > 0$ by 't Hooft
vertex, which for one flavor is just an additive mass shift
\cite{Creutz:2007yr}.  Unlike the case with more flavors, this
expectation cannot be regarded as a spontaneous symmetry breaking
since there is no chiral symmetry to break.  Any regulator that
preserves a remnant of chiral symmetry must inevitably fail
\cite{Creutz:2008nk}.  Note also that there is no longer the necessity
of a first order phase transition at $\Theta=\pi$.  It has been argued
\cite{Creutz:2006ts} that for finite quark mass such a transition can
occur if the mass is sufficiently negative, but the region around
vanishing mass has no distinguishing structure.

One feature of one-flavor QCD is that the renormalization of the quark
mass is not multiplicative when non-perturbative effects are taken
into account.  The additive mass shift is generally scheme dependent
since the details of the instanton effects depend on scale.  This is
the basic reason that a massless up quark is not a possible solution
to the strong CP problem \cite{Creutz:2003xc}.

Because of this shift, the conventional variables $\Theta$ and $m$ are
singular coordinates for the one-flavor theory.  A cleaner set of
variables would be the coefficients of the two possible mass terms
$\overline\psi \psi$ and $i\overline\psi \gamma_5\psi$ appearing in
the Lagrangian.  The ambiguity in the quark mass is tied to rough
gauge configurations with ambiguous winding number.  This applies even
to the formally elegant overlap operator; when rough gauge fields are
present, the existence of a zero mode can depend on the detailed
operator chosen to project onto the overlap circle.  Smoothness
conditions imposed on the gauge fields to remove this ambiguity appear
to conflict with fundamental principles, such as reflection positivity
\cite{Creutz:2004ir}.

\section{Varying $N_f$}
\label{nondegenerate}

The $Z_{N_f}$ symmetry discussed here is a property of the fermion
determinant and is independent of the gauge field dynamics.  In Monte
Carlo simulation language, this symmetry appears configuration by
configuration.  With $N_f$ flavors, we always have $|D|=|e^{2\pi
i/n_f} D|$ for any gauge field.  This discrete chiral symmetry is
inherently discontinuous in $N_f$.  This non-continuity lies at the
heart of the controversy over the rooted staggered quark approximation
to lattice gauge theory.  The details of this issue are extensively
discussed elsewhere \cite{Creutz:2008nk}, but the essence is that the
four species inherent with staggered quarks give rise to an unphysical
extra $Z_4$ which current algorithms do not remove.

\begin{figure*}
\centering \includegraphics[width=2.5in]{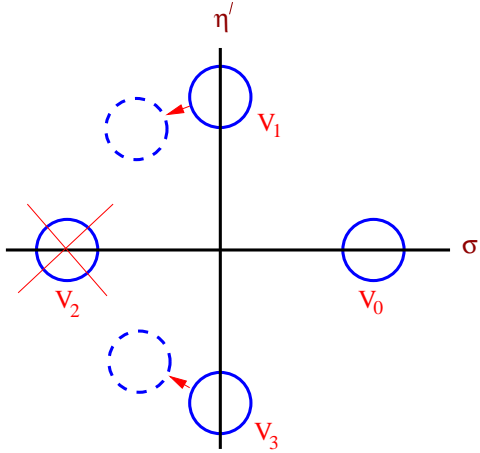}
\caption{\label{potential4} 
As one takes the four flavor theory and increases the mass of one
quark, one of the four original minima of the effective potential
should disappear while the others rearrange to give the final three
fold symmetry.
}
\end{figure*}

It is possible to interpolate between various numbers of flavors by
adjusting the quark masses.  Construct an $SU(N)$ valued effective
field $\Sigma=\exp(i\pi_\alpha\lambda_\alpha)$.  If, for instance, we
give one flavor a large mass, this will drive one component of
$\Sigma$ to unity
\begin{equation}
\Sigma_{N_f} \longrightarrow
\pmatrix{\Sigma_{N_f-1} & 0\cr 0& 1},
\end{equation}
leading one to the above discussion with one less flavor.  A
complication arises since the breaking of the $SU(N_f)$ flavor
symmetry by a non-singlet mass term allows mixing of the $\eta^\prime$
field with the flavored analog of the $\eta$.  As the heavier quark
mass increases we should adjust what is meant by $\eta^\prime$, but
qualitatively the transformation from, say, four to three flavors
should look something like what is sketched in Fig.~\ref{potential4},
where one of the minima moves up to disappear and the others rearrange.

\begin{figure*}
\centering
\includegraphics[width=2.5in]{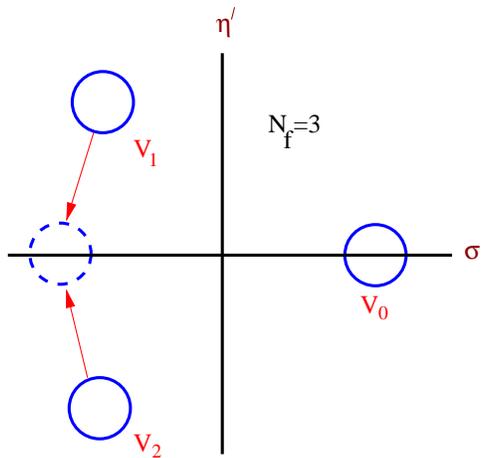}
\caption{\label{potential5}
Going from three to two flavors by increasing the mass of the strange
quark should result in two of the minima of the effective potential
merging into one.
}
\end{figure*}

Continuing down from an odd number of flavors to an even number, then
the two degenerate minima at negative mass should merge, as sketched
in Fig.~\ref{potential5}.  For arbitrary masses the situation becomes
increasingly complicated.  In Ref.~\cite{Creutz:2003xu} the expected
phase diagram is mapped out for the case of three flavors with
arbitrary real masses.  There it is shown that there are large regions
both with and without a first order transition at $\Theta=\pi$.

The first order transition at $\Theta=\pi$ remains robust as long as
multiple lightest quarks are degenerate and their masses remain in the
regime where chiral expansions make sense.  Regardless of any heavier
quarks, the above symmetry arguments hold for the light quarks when
the heavier masses are held fixed.  Only when the lightest quark is
non-degenerate can a gap appear separating the region of spontaneous
CP violation at negative quark mass from zero mass.  The size of this
gap is controlled by the heavier quark masses.

\section{Summary}
\label{summary}

I have discussed how the anomalous breaking of the classical $U(1)$
axial symmetry in QCD interplays with the spontaneously broken
flavored axial symmetries.  With $N_f$ massless quarks a flavor
singlet discrete $Z_{N_f}$ chiral symmetry is left behind.  This
provides an intuitive interpretation of the strong CP violating angle
$\Theta$ in terms of effective meson fields.  As a consequence, a
first order transition is generally expected at $\Theta=\pi$ and $m\ne
0$.  This is quite robust and can only be avoided if one quark is
considerably lighter than the others.  At $\Theta=\pi$ the QCD
Lagrangian is CP invariant; so, the transition represents a
spontaneous breaking of this discrete symmetry.  The physics of the
anomaly indicates that the signs of quark masses can be significant,
something that is not naturally interpreted perturbatively.  In the
special case of one flavor, all chiral symmetry is lost.  One
consequence is that the mass of a non-degenerate light quark is
unprotected from an additive and scheme dependent renormalization.
Furthermore, any proposed regulator that maintains an exact chiral
symmetry for the one flavor case must fail.

\section*{Acknowledgments}
This manuscript has been authored under contract number
DE-AC02-98CH10886 with the U.S.~Department of Energy.  Accordingly,
the U.S. Government retains a non-exclusive, royalty-free license to
publish or reproduce the published form of this contribution, or allow
others to do so, for U.S.~Government purposes.

\end{document}